\documentclass{aa}

\usepackage{graphicx}
\usepackage{natbib}
\usepackage{txfonts}
\usepackage{setspace}

\begin{document}

\title{On the globular cluster formation history of NGC 5128}

\author{S. Kaviraj  \inst{1}     \and
        I. Ferreras \inst{1,2,3} \and
        S.-J. Yoon  \inst{1}     \and 
        S. K. Yi    \inst{1}}

\institute{Department of Physics, University of Oxford, Keble Road, Oxford OX1 3RH \and
           Department of Physics, Institute of Astronomy, 
           ETH Hoenggerberg HPF D8, 8093 Zurich, Switzerland \and
           Department of Physics and Astronomy, University College London, Gower Street, London WC1E 6BT}

\offprints{Sugata Kaviraj, \email{skaviraj@astro.ox.ac.uk}}

\date{Accepted (in press)}

\abstract{We deduce the globular cluster formation history of the nearby
elliptical galaxy, NGC 5128, by using a chemical enrichment model to
accurately reproduce its observed metallicity distribution
function (MDF). We derive the observed MDF using recently obtained
$U$ and $B$ photometry of the NGC 5128 GC system, with $(U-B)$
used as the metallicity indicator. Our results indicate that the GC system
in this galaxy could be the product of two major GC formation
episodes. The initial formation episode occured 11-12 Gyrs ago creating
65-75 percent of the mass in the GC system. This was followed by a second late
formation episode which peaked 2-4 Gyrs ago and produced the
remaining 25-35 percent of GC mass.

}

\maketitle

\section{Introduction}
Globular cluster (GC) systems have been popular tools in
deciphering the star formation histories (SFHs) of galaxies. Their
ubiquitous presence in galaxies of all morphological types,
coupled with evidence that their creation seems to accompany major
star formation episodes \cite[e.g.][]{LR99} makes them useful
tracers of galactic evolution
\cite[e.g.][]{KP98,VDB2000,Yoon2002}. In addition, GC systems are
well represented by simple stellar populations (SSPs) with stars
of the same age and chemical composition, which makes them easy to
study using SSP models \cite[e.g.][]{Yi2004}.

The exact formation mechanism of GCs has been the subject of much
recent debate. Many models for GC formation have been proposed
including gaseous mergers \citep{Ashman92}, in situ formation
\citep[e.g.][]{Harris95}, multiphase collapse \citep{Forbes97},
dissipationless hierarchical merging
\citep[e.g.][]{Cote98,Cote2000,Cote2002} and hierarchical clustering
\citep{Beasley2003}. While none of these models can be
conclusively excluded, the widespread discovery of multimodal
metallicity distributions in GC populations effectively rules out
an extreme version of the monolithic collapse scenario for their
formation \citep[see e.g.][]{Forbes97}. Furthermore, the correlation between
the mean metallicity of GCs and galaxy luminosity indicates that chemical
enrichment of the GC system is intimately linked to the evolution
of the host galaxy
\citep{Forbes96,Durrell96,Forbes2001,Cote2000,Jordan2004}. An elegant
review of the GC formation models mentioned above can be found in \cite{West2004}.

The object of this study is NGC 5128, the giant elliptical galaxy
in the nearby Centaurus Group \citep[see][for a comprehenive
review of NGC 5128]{I98}. Located at a distance of approximately 3.6 Mpc
\citep[e.g.][]{Soria96}, NGC 5128 is the closest giant elliptical
system with an estimated GC population of $1550 \pm 350$
\citep{Harris84}. It has been widely studied, not only because of
its proximity and relative brightness, but also because it
displays unusual physical features which suggest that this galaxy
is a post-merger remnant. A prominent dust lane containing
young stars and HII regions \citep[e.g.][]{Unger2000,Wild2000} and
a series of optical shells \citep{Malin83}, which have HI
\citep{Schiminovich94} and molecular CO \citep{Charmandaris2000}
gas associated with them, are considered strong evidence that NGC
5128 underwent a recent merger event within the last $10^9$ years.
Recently, \citet{Rejkuba2004} suggested that star formation may have
stopped as recently as 2 Myr ago in the north-eastern shell of NGC
5128. 

In a series of major works, \citet{Harris99}, \citet{Harris2000} and
\citet{Harris2002} performed a comprehensive
study of the metallicity distribution of stars in the inner and outer
halo of NGC 5128. In this study we deduce the formation history of the GC system of
NGC 5128 by accurately reproducing its observed metallicity
distribution function (MDF) using a chemical enrichment model. We
derive the observed MDF of this elliptical galaxy using recently
obtained $U$ and $B$ photometry of 210 clusters in its GC system
\citep{Peng2004a}. The integrated $(U-B)$ colour is used as the
metallicity indicator because it is sensitive to
metallicity via the opacity effect but relatively insensitive to
the effective main sequence turn-off temperature
($T_{eff}$) and therefore to age when
$T_{eff} \sim$ 7000-12000 K \citep{Yi2004}. Similar techniques using
$U$ band colours have been used by \citet{rejkuba2001} and
\citet{jordan2002} who used $(U-V)$ and \emph{Hubble Space
Telescope} WFPC2 $(F336W-F547M)$, respectively. Although these colours
are substantially better metallicity
indicators than the previously used $(V-I)$ or $(B-V)$, they still change gradually with age and thus are not
as effective as $(U-B)$ in determining metallicity \citep{Yi2004}. Our work
extends previous studies of NGC 5128 in a number of ways. The
large number of confirmed GCs with $U$ band photometry makes it
possible to derive statistically significant results. In addition,
the relative robustness of $(U-B)$ as a metallicity indicator,
compared to other optical colours, makes chemical enrichment
an effective approach in modelling the
formation history of the NGC 5128 GC system. The use of a
consistent chemical enrichment code means that we can effectively
transform metallicities into ages.

We begin by briefly checking that, as might be expected from previous theoretical
and observational results, a single starburst followed by passive
evolution (a monolithic scenario) is incapable of reproducing the
observed MDF of the NGC 5128 GCs. Performing this check is
not a redundant exercise because our modelling essentially yields
\emph{relative likelihoods} for various models to fit the observed
MDF of NGC 5128. It is therefore instructive to compare the quality of
fit between monolithic and non-monolithic scenarios. The main
thrust of the paper, however, is to explore a \emph{double starburst
scenario}, analysing positions, timescales and relative strengths
of the two star formation episodes that best explain the MDF of
the NGC 5128 GC system. We provide a coherent picture of the formation
history of NGC 5128 GCs based on double starburst scenarios
that give excellent fits to the observed MDF and show that our results
are consistent with the spectroscopic study of \citet{Peng2004b}, who
use the age-sensitive $H_{\beta}$ index to age-date the GCs in
this dataset.

\section{Derivation of the observed NGC 5128 MDF}
We derive the observed MDF for the NGC 5128 GC system by using
photometric data in the $U$ and $B$ bands. We compute the
metallicities by overlaying a model $(U-B)$ vs. $(B-V)$ grid on
the GC data, with $(U-B)$ used as the metallicity indicator.

\begin{figure}
\begin{center}
\includegraphics[width=3.4in]{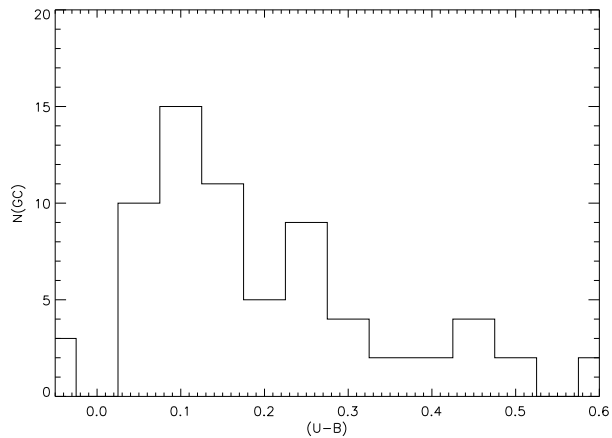}
\caption{$(U-B)$ colour distribution derived from final sample of GCs
  used in this study.}
\label{fig:ub_distribution}
\end{center}
\end{figure}

The metallicity of each GC is computed by identifying its two
bounding iso-metallicity curves, followed by linear interpolation
between them. We do not use the same iso-metallicity curves to
compute the metallicity error from the $(U-B)$ errors. Instead we
treat each error limit as a separate \emph{point} on the grid and
repeat the process described above. Given the irregular spacing
and pseudo-horizontal nature of the model iso-metallicity curves
(see Fig. 7 in Yi et al. 2004), this results in a more accurate
approximation of the metallicity errors for each GC. It
should be noted that $(U-B)$ colours (especially redder colours) are
affected by the age-metallicity degeneracy. The colour lying in a
region of large degeneracy induces a larger metallicity error,
since the corresponding metallicity measurement becomes more uncertain. We also note that
\citet{Yi2004} did not attempt to accurately derive the ages and
metallicities of individual GCs in this dataset. The age and metallicity 
derivations given in \citet{Yi2004} are rough estimates derived
from average values of $(B-V)$ and $(U-B)$ in broad bins. The method
we use here provides far superior quantification of the
metallicity by using a finely interpolated $(U-B)$ grid.

\section{GC selection}
Our photometric data on NGC 5128 contains 210 GCs, some of which
suffer from large uncertainties in the $U$ and $B$ magnitudes. In
this study we applied a cut of 0.1 mag to the uncertainty in
$(U-B)$, retaining 69 GCs out of the original sample of 210. The
reasons for applying a rather stringent magnitude error cut are
twofold. Firstly, some GCs have extremely large photometric errors.
By using GCs with small error bars we are able to better quantify
the metallicity of the GCs that we eventually use along with their
associated errors. This in turn results in more constrained
estimates for the characteristics of the star formation episodes.
Secondly, having performed this magnitude cut we find that the MDF for
the entire sample of 210 GCs is
very similar to our reduced sample; a Kolmogorov-Smirnov (KS) test
\citep[see e.g.][]{Wall96}
between the entire sample and the distribution derived after the
uncertainty cut yields a KS probability of 42 percent, indicating that
the two samples are consistent at the 95 percent confidence
level. 15 percent of the final sample of 69 GCs lie within a projected radius of
6' from the centre of the galaxy (NGC 5128 has an apparent size of 18'
$\times$ 14') compared to 25 percent of the original sample. 56
percent of the final sample lies within 10' compared to 59 percent of
the original sample. The faint end of the GCLF of the final sample
lies at $V=19.7$ mag, while the faint-magnitude cuts are $20.6$ and
$20.4$ in the $U$ and $B$ bands respectively.

\begin{figure}[!ht]
\begin{center}
\includegraphics[width=3.4in]{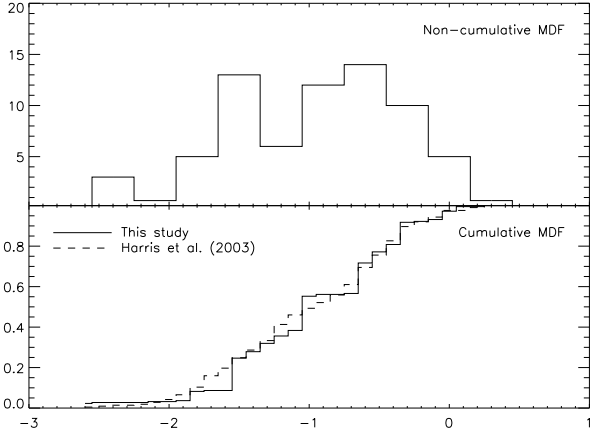}
\caption{TOP PANEL: Metallicity
distribution derived from $(U-B)$ colours of 69 GCs which have
$(U-B)$ photometric errors less than 0.1 mag. All photometry is
corrected for Galactic extinction. BOTTOM PANEL: Comparison between
the derived metallicity distribution (shown in top panel) and that
derived by Harris et al. (2003) using the metallicity sensitive
$(C-T_1)$ index.}
\label{fig:observed_mdf}
\end{center}
\end{figure}

Our aim is to maximise the \emph{accuracy} of our analysis. The
magnitude cut we employ increases the robustness of our final
solutions, without transforming the original distribution
significantly. Fig. \ref{fig:ub_distribution} presents the $(U-B)$
colour distribution derived from our final sample of clusters and
Fig. \ref{fig:observed_mdf} presents the
metallicity distribution derived from the $(U-B)$ colours of our
final sample of clusters. We note that our derived MDF is
consistent with the NGC 5128 MDF derived using the ($C-T_1$) index
for 205 clusters in \citet{Harris2003}. This comparison is shown in
the bottom panel of Fig. 
\ref{fig:observed_mdf}.

\section{Star formation and gas infall prescriptions}
Feedback processes from star formation remain poorly understood.
Supernovae typically inject large amounts of energy into the
inter-stellar medium. The energy deposited can be either
kinetic or thermal and the exact description of these processes
remains the subject of much debate \citep[see
e.g.][]{Katz92,Mihos94,Yepes97,Ferreras2002,Silk2003}. Such
feedback processes can interrupt or accelerate star formation, so
the simple coupling between star formation and the gas density
implied by, for example, a Schmidt law \citep{Schmidt59} may not
hold. In this model, we decouple the star formation rate from the
gas density in the system to qualitatively account for such
feedback processes from star formation. We take the star formation
rate to be a sequence of gaussian functions independent of the gas
density

\begin{equation}
\Psi(t)=\sum_n \frac{f_n}{\sqrt{2\pi}\sigma_n} \exp
{\frac{-(t-\tau_n)^2}{2\sigma^2_n}}.
\label{star_formation_rate_equation}
\end{equation}

In Eq. \ref{star_formation_rate_equation} $f_n$ is the
normalisation amplitude, $\sigma_n$ the temporal spread and
$\tau_n$ the centre-point of starburst $n$. With the gaussian
starbursts normalised as shown above, we keep the total area under
$\Psi(t)$ and hence the total stellar mass equal to 1 by ensuring
that $\sum_n f_n=1$. We checked that our results do not
depend strongly on our choice of mathematical function by
repeating our analysis using lorentzian starbursts, and find that
our conclusions remain unchanged.

We assume that the system begins with no gas and use an
exponentially decaying gas infall rate

\begin{equation}
g_{in}(t)=\frac{\eta}{\tau_{in}}\exp\Big({\frac{-t}{\tau_{in}}}\Big).
\end{equation} Here $\tau_{in}$ is the infall timescale, and $\eta$ is a parameter
that controls the final gas fraction in the system and is defined
below. Noting that,

\begin{equation}
\int^{t_U}_{0}
g_{in}(t)dt=\eta\Big[1-\exp\Big(\frac{-t_U}{\tau_{in}}\Big)\Big]\approx\eta,
\end{equation}
where $t_U$ is the age of the universe in Gyrs, we can force the
system to have a final gas fraction $g_{frac}$ by choosing
\begin{equation}
g_{frac}=\frac{\eta-1}{\eta},
\end{equation}
so that,
\begin{equation}
\eta=\frac{1}{1-g_{frac}}.
\end{equation}
We fix the infall timescale $\tau_{in}$ to 5 Gyrs, which leaves the
following $3n$ free parameters in the model for an $n$ starburst
scenario
\begin{eqnarray*}
 &(f_1,\ldots,f_{n-2},\tau_1,\ldots,\tau_{n-1},\sigma_1,\ldots,\sigma_{n-1},g_{frac}).&
\end{eqnarray*} We note that due to the decoupling of the star formation rate from
gas density, changing the value of the $\tau_{in}$ has no impact
on our subsequent analysis. Essentially, the behaviour of the
system remains unchanged as long as there is a reservoir of gas
available to make stars.

Given the prescriptions given above for the star formation and gas
infall rates, we compute the metallicity evolution of the
GCs following the standard set of chemical
enrichment equations as described in \citet{Ferreras2000b}.
We use the yields of \citet{Thielemann96} for stellar
masses $M_\star>10M_\odot$ and \citet{vandenHoek97}
for lower mass stars. For a given set of parameters we
integrate the equations to allow us to compute the MDF, that is then
compared with the observed MDF via a KS test.

\section{Derivation of best-fit parameters}
We begin our analysis by comparing a single
starburst scenario, which describes a monolithically evolving
system, to the more realistic ones that
underpin this study. We then refine our analysis by adding another
starburst and perform a detailed study of the resulting six dimensional
parameter space. We also check that small perturbations to the two major
star formation episodes do not affect the best-fit values for our
parameters. This sequence is repeated for four final gas fractions
- $g_{frac}=0.1,0.2, 0.35, 0.5$. Since we are modelling an
elliptical galaxy, the value of $g_{frac}$ is likely to be low,
probably somewhere in the region $0.1-0.35$. In a recent work,
\citet{Sanderson2003} finds the (hot) gas fraction in ellipticals
within $R_{200}$ to be approximately $0.1$.

We compare the MDF predicted by our model to the observed GC MDF using
the KS test. The
\emph{best-fit} values of parameters are calculated by
marginalising their KS probabilities, which involves summing
(integrating) out the probability dependence of all other
parameters, leaving a probability distribution for each parameter that
is independent of all others in the model. For example, in a
double starburst scenario, the marginalised probability
distribution for the parameter $\tau_1$ is calculated as

\begin{equation}
P(\tau_1)=\varsigma
\sum_{i,j,k}P(\tau_1,\tau_{2i},\sigma_{1j},\sigma_{2k}),
\end{equation}
where $\varsigma$ is a normalising constant which ensures that
$P(\tau_1)$ integrates to 1. This probability distribution will
not, in general, be symmetric and therefore should not be
approximated by a gaussian.

We take the best-fit value of a parameter as the value at which
its marginalised probability function peaks. In addition, we
\emph{define} errors for these parameters superficially
similar to the \emph{one-sigma} errors frequently used for
gaussian probability distributions. If $X$ is the most likely
value of a parameter $x$, then we define the positive error $x_{+}$ as
\begin{equation}
\int_{X}^{x_{+}} P(x)dx=0.34 \int_{X}^\infty P(x)dx,
\end{equation}
and similarly the negative error, $x_{-}$, as
\begin{equation}
\int_{x_{-}}^{X} P(x)dx=0.34 \int_{-\infty}^X P(x)dx.
\end{equation}

Finally, given the fact that each GC has a metallicity error
associated with it, we check that the positions of the most likely
values for our model parameters are robust with respect to these
errors. We perform this check by adding random noise to our
observed MDF and producing KS values for these \emph{noisy} MDFs.
The noise is added by running through each GC and changing its
metallicity to one of its error points \emph{randomly}. This
process is repeated a hundred times to generate a hundred noisy
MDFs. We have checked that the KS values do not deviate
significantly from the fiducial value and find that the KS values
are suitably robust (within 6 percent of the original values) when
the fiducial MDF is perturbed in this way. This is mainly a result
of choosing GCs with small photometric errors.

\section{Results and analysis}
\subsection{Double starburst scenarios}

For a single starburst scenario, the maximum KS probability we
achieve in our model is around 2 percent, significantly lower than
the double starburst scenarios we explore later. The most favoured
values of $\tau$ and $\sigma$ for the single starburst scenario are
$~1$ Gyr and $1.35$ respectively. A single starburst
model fails to reproduce, with any acceptable
accuracy, the observed MDF of NGC 5128.

With two starbursts we obtain a six dimensional
parameter space with the star formation rate as the sum of two
gaussians

\begin{equation}
\Psi(t)=\frac{f_1}{\sqrt{2\pi}\sigma_1} \exp
{\frac{-(t-\tau_1)^2}{2\sigma^2_1}}+\frac{1-f_1}{\sqrt{2\pi}\sigma_2}
\exp {\frac{-(t-\tau_2)^2}{2\sigma^2_2}}.
\end{equation} Table 1 shows the best-fit values for the starburst parameters ($f_1$, $f_2$,
$\tau_1$, $\tau_2$, $\sigma_1$, $\sigma_2$) for final gas
fractions of $0.5$ to $0.1$. Note that in Table 1,
  Subscript '1' indicates the initial (early) starburst and Subscript
  2 indicates the second (late) starburst.

\begin{table*}
\begin{center}
\begin{minipage}{126mm}
\caption{\small{Best-fit values and errors for the double
starburst scenarios with final gas fractions in the range
$0.1-0.5$. Note that Subscript '1' indicates the initial (early) starburst and Subscript
  2 the second (late) one.}}
\begin{tabular}{c|c|c|c|c|c}
    \multicolumn{6}{c}{Final gas fraction = 0.5}\\
    $(f_1,f_2)$ & $(0.5,0.5)$ & $(0.6,0.4)$ & $(0.65,0.35)$ & $(0.8,0.2)$ & $(0.9,0.1)$\\
    \hline \hline
              &  &  &  &  \\
$\tau_1$      & $0.72^{+0.12}_{-0.31}$ & $1.00^{+0.09}_{-0.14}$ & $0.51^{+0.18}_{-0.02}$  & $1.01^{+0.13}_{-0.04}$ & $1.00^{+0.14}_{-0.21}$\\
              &  &  &  &  & \\
$\tau_2$      & $5.00^{+0.22}_{-0.26}$ & $9.99^{+0.46}_{-0.26}$ & $10.00^{+0.44}_{-1.00}$ & $11.90^{+0.13}_{-0.96}$ & $11.10^{+0.25}_{-0.95}$\\
              &  &  &  &  \\
$\sigma_1$    & $0.70^{+0.01}_{-0.12}$ & $0.70^{+0.15}_{-0.03}$ & $1.00^{+0.12}_{-0.05}$ & $1.02^{+0.23}_{-0.08}$ & $2.00^{+0.13}_{-0.09}$\\
              &  &  &  &  & \\
$\sigma_2$    & $3.55^{+0.21}_{-0.27}$ & $3.58^{+0.24}_{-0.17}$ & $1.99^{+0.80}_{-0.25}$ & $0.60^{+1.02}_{-0.02}$ & $1.49^{+0.17}_{-0.23}$ \\
              &  &  &  &  & \\
KS prob       & 0.57 & 0.43 & 0.61 & 0.13 & 0.06\\
              &  &  &  &  & \\

    \hline \hline
    \multicolumn{6}{c}{}\\
    \multicolumn{6}{c}{Final gas fraction = 0.35}\\
    $(f_1,f_2)$ & $(0.5,0.5)$ & $(0.6,0.4)$ & $(0.65,0.35)$ & $(0.8,0.2)$ & $(0.9,0.1)$\\
    \hline \hline
              &  &  &  &  \\
$\tau_1$      & $1.00^{+0.20}_{-0.09}$ & $1.00^{+0.20}_{-0.09}$ & $1.00^{+0.20}_{-0.06}$  & $1.00^{+0.23}_{-0.06}$ & $1.00^{+0.20}_{-0.09}$\\
              &  &  &  &  & \\
$\tau_2$      & $11.40^{+0.07}_{-0.54}$ & $9.00^{+0.07}_{-0.93}$ & $12.00^{+0.07}_{-0.93}$ & $9.00^{+0.83}_{-0.88}$ & $12.10^{+0.62}_{-0.92}$\\
              &  &  &  &  \\
$\sigma_1$    & $3.02^{+0.38}_{-0.93}$ & $1.10^{+0.20}_{-0.09}$ & $1.13^{+0.19}_{-0.09}$ & $1.99^{+0.31}_{-0.22}$ & $2.99^{+0.28}_{-0.42}$\\
              &  &  &  &  & \\
$\sigma_2$    & $0.70^{+0.16}_{-0.04}$ & $3.00^{+0.36}_{-0.28}$ & $0.99^{+0.48}_{-0.15}$ & $3.08^{+0.36}_{-0.55}$ & $3.05^{+0.36}_{-0.73}$ \\
              &  &  &  &  & \\
KS prob       & 0.61 & 0.47 & 0.73 & 0.05 & 0.03\\
              &  &  &  &  & \\

    \hline \hline
    \multicolumn{6}{c}{}\\
    \multicolumn{6}{c}{Final gas fraction = 0.2}\\
    $(f_1,f_2)$ & $(0.5,0.5)$ & $(0.6,0.4)$ & $(0.65,0.35)$ & $(0.8,0.2)$ & $(0.9,0.1)$\\
    \hline \hline
              &  &  &  &  \\
$\tau_1$      & $0.99^{+0.15}_{-0.04}$  & $1.00^{+0.11}_{-0.06}$ & $1.01^{+0.13}_{-0.04}$  & $0.99^{+0.15}_{-0.04}$ & $1.00^{+0.16}_{-0.04}$\\
              &  &  &  &  & \\
$\tau_2$      & $10.50^{+0.33}_{-0.88}$ & $11.45^{+0.19}_{-0.48}$ & $11.20^{+0.15}_{-0.61}$ & $10.90^{+0.24}_{-1.32}$ & $10.70^{+0.24}_{-1.46}$\\
              &  &  &  &  \\
$\sigma_1$    & $0.76^{+0.19}_{-0.05}$  & $0.99^{+0.11}_{-0.06}$ & $1.00^{+0.14}_{-0.07}$ & $2.00^{+0.19}_{-0.15}$ & $2.99^{+0.14}_{-0.61}$\\
              &  &  &  &  & \\
$\sigma_2$    & $3.50^{+0.23}_{-0.18}$  & $1.74^{+0.31}_{-0.27}$ & $1.00^{+0.37}_{-0.10}$ & $2.00^{+0.38}_{-0.22}$ & $2.50^{+0.66}_{-0.20}$ \\
              &  &  &  &  & \\
KS prob       & 0.49 & 0.43 & 0.44 & 0.04 & 0.02\\
              &  &  &  &  & \\

    \hline \hline
    \multicolumn{6}{c}{}\\
    \multicolumn{6}{c}{Final gas fraction = 0.1}\\
    $(f_1,f_2)$ & $(0.5,0.5)$ & $(0.6,0.4)$ & $(0.65,0.35)$ & $(0.8,0.2)$ & $(0.9,0.1)$\\
    \hline \hline
              &  &  &  &  \\
$\tau_1$      & $1.12^{+0.11}_{-0.12}$ & $0.91^{+0.12}_{-0.67}$ & $1.45^{+0.20}_{-0.12}$  & $1.20^{+0.18}_{-0.12}$ & $1.00^{+0.18}_{-0.34}$\\
              &  &  &  &  & \\
$\tau_2$      & $11.20^{+0.15}_{-0.78}$ & $12.06^{+0.15}_{-0.28}$ & $11.40^{+0.11}_{-0.87}$ & $12.00^{+0.56}_{-0.78}$ & $10.89^{+0.14}_{-0.80}$\\
              &  &  &  &  \\
$\sigma_1$    & $0.72^{+0.29}_{-0.18}$ & $0.99^{+0.12}_{-0.03}$ & $1.00^{+0.11}_{-0.05}$ & $3.00^{+0.29}_{-0.06}$ & $2.50^{+0.10}_{-0.35}$\\
              &  &  &  &  & \\
$\sigma_2$    & $3.53^{+0.28}_{-0.58}$ & $3.46^{+0.23}_{-0.17}$ & $2.00^{+0.17}_{-0.30}$ & $1.00^{+0.36}_{-0.12}$ & $3.00^{+0.56}_{-0.18}$ \\
              &  &  &  &  & \\
KS prob       & 0.14 & 0.22 & 0.03 & 0.03 & 0.02\\
              &  &  &  &  & \\

    \hline

\end{tabular}
\end{minipage}
\end{center}
\label{tab:marginalised_values}
\end{table*}

\begin{figure}
\begin{center}
\includegraphics[width=3.4in]{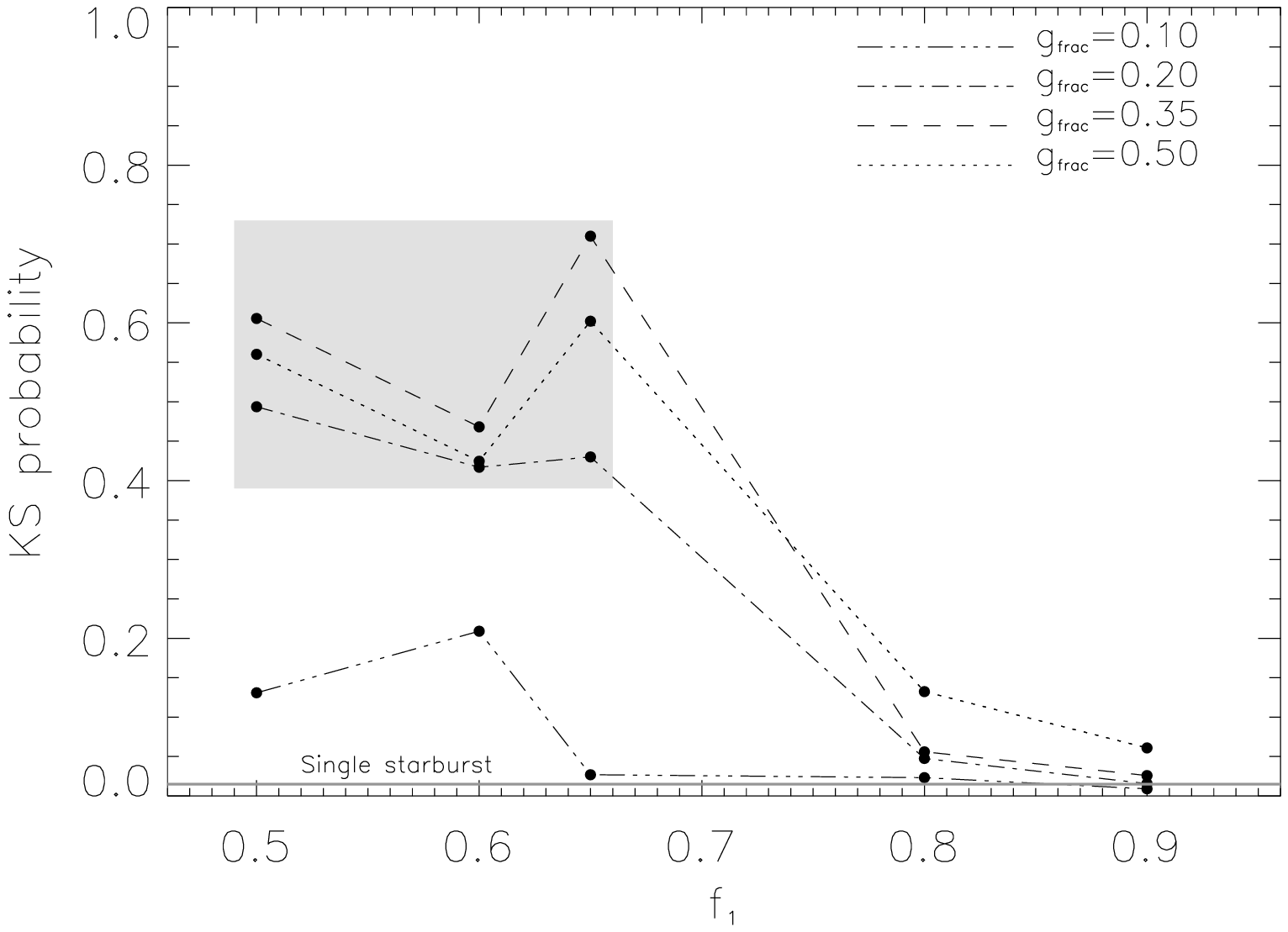}
\caption[Best KS probabilities for double starburst
scenarios]{\small{Best KS probabilities for various values of
$f_1$ and for four final gas fractions $g_{frac}$. The shaded
region shows a set of degenerate models that have high KS
probabilities and therefore cannot be discounted completely. The
KS probability for a single starburst scenario is shown by the
solid line at 2 percent.}} \label{fig:relative_KS_values}
\end{center}
\end{figure}

\begin{figure}
\begin{center}
\includegraphics[width=3.4in]{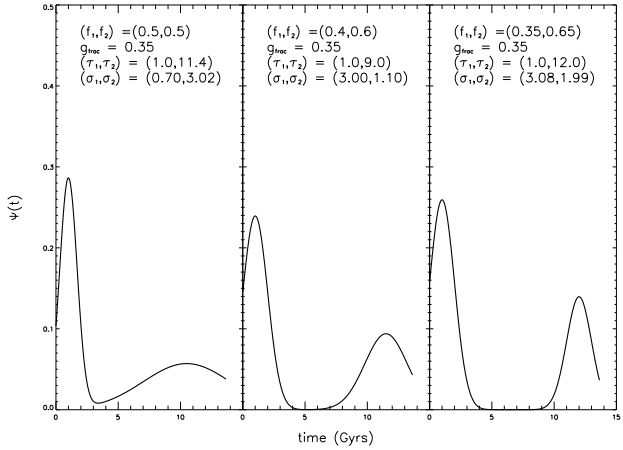}
\caption[Star formation histories in best-fit double starburst
models]{\small{Three GC formation histories corresponding to the
best KS probabilities shown in the shaded region in Fig.
\ref{fig:relative_KS_values}.}} \label{fig:best_sfr}
\end{center}
\end{figure}

In Fig. \ref{fig:relative_KS_values}, we present the best KS
values produced in the double starburst models. The single
starburst case is shown, for comparison, by the solid line at 2
percent. In Fig. \ref{fig:best_sfr} we present three GC
  formation histories which yield the highest KS
probabilities (see shaded region in Fig.
\ref{fig:relative_KS_values}) by showing the formation histories for favoured values
($f_1,f_2$) in a system with a final gas fraction of 0.35.

\begin{figure}
\begin{center}
\includegraphics[width=3.4in]{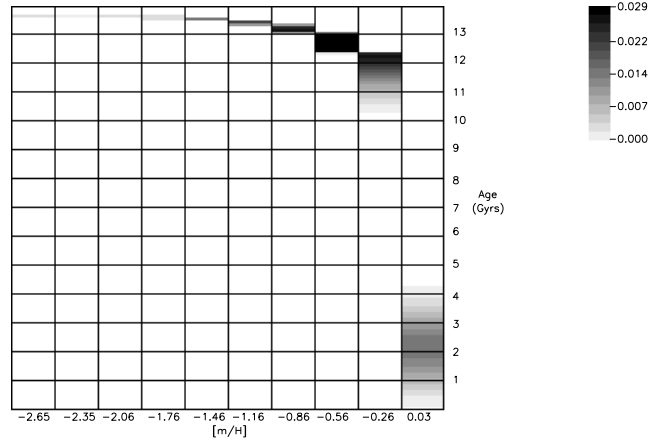}
\caption{Age-metallicity grid showing the fraction of
stellar mass predicted by the favoured model as a
    function of age and metallicity. The key indicates stellar
    fractions corresponding to the levels of shading in the plot.}
\label{fig:synthetic_sfhtab}
\end{center}
\end{figure}

\begin{figure}
\begin{center}
\includegraphics[width=3.4in]{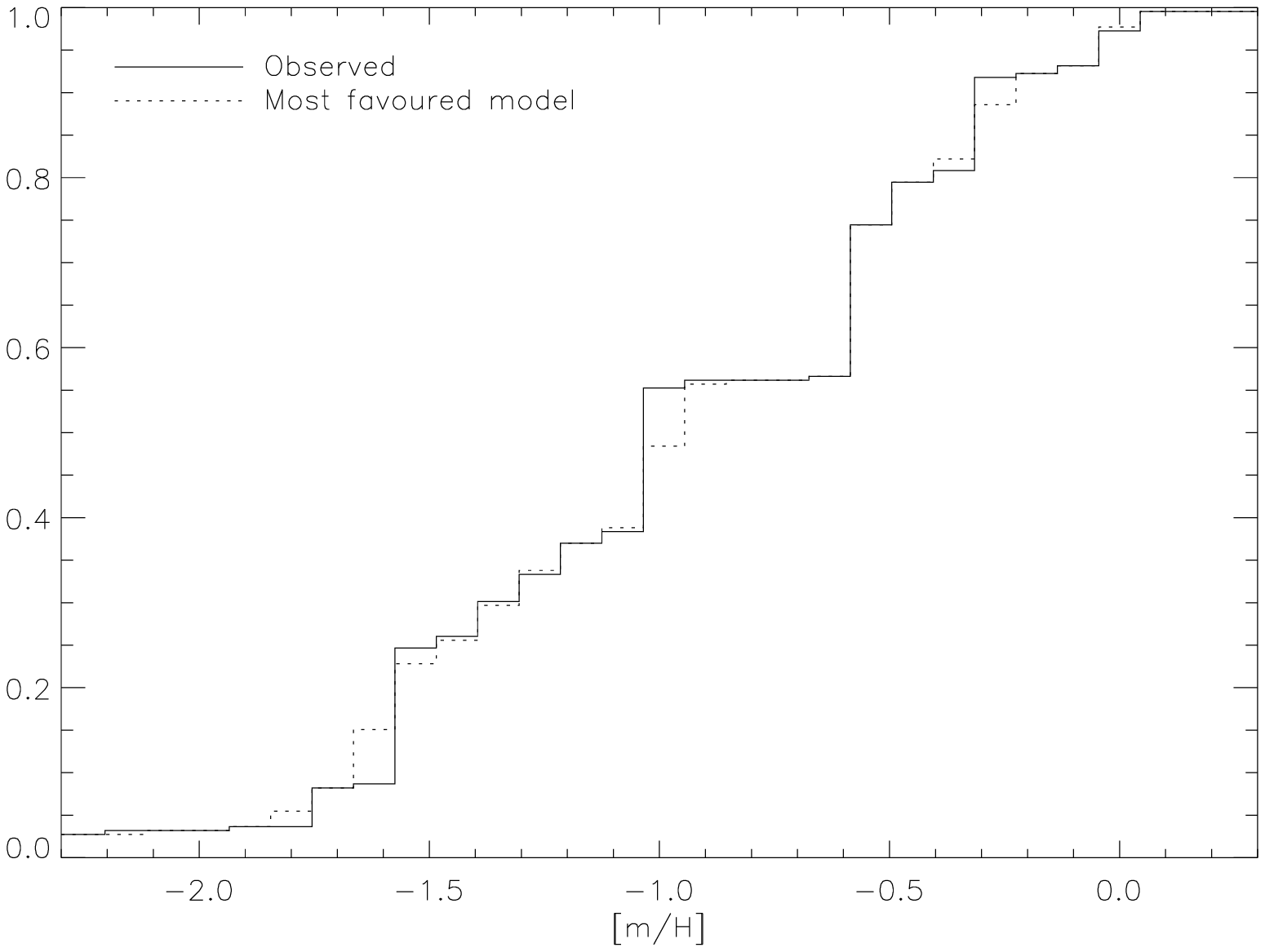}
\caption{Comparison between observed MDF and that predicted by the
  favoured model (which gives the highest KS probabillity).}
\label{fig:mdf_match}
\end{center}
\end{figure}

Fig. \ref{fig:relative_KS_values} suggests that an intermediate
gas fraction of $0.35$ with $(f_1,f_2)=(0.35,0.65)$ is the
\emph{most favoured model}, so that roughly one-third of the mass in
the GC system is predicted to have been formed in a late starburst
centred around 2 Gyrs ago. Fig. \ref{fig:synthetic_sfhtab} shows
the \emph{mass fractions} formed at various ages and metallicities in
this model, and Fig. \ref{fig:mdf_match} compares the MDF predicted
by this model to the observed MDF.

Although the favoured model provides the best-fit, the shaded region in Fig.
\ref{fig:relative_KS_values} shows a set of degenerate solutions
consisting of models producing high KS values, all of which
reasonably reproduce the observed NGC 5128 MDF and therefore
cannot be completely discounted. Changing the relative sizes of the starbursts
even further results in the KS probability declining rapidly. This
is clearly expected, since in the limit $f_1 \rightarrow 0$, we
recover a single starburst, already established as
being incapable of reproducing the observed NGC 5128 MDF. While the position of the first
  starburst is fairly robust in the best-fit models, the second
  starburst tends to vary, so that we are left with a
dispersion of $\sim$ 2 Gyrs in the second
starburst's position. This reflects the fact that uncertainties
in colours and in the age-metallicity degeneracy (discussed in
Sect. 2) induce uncertainties
in the metallicity derivations for our GCs, which in turn limits our resolution
of the position of the late GC formation episode.
Averaging over all the best-fit models (shown in the shaded region
in Fig. \ref{fig:relative_KS_values}) and taking the one-sigma errors
into account for the parameters given in Table 1, our results therefore indicate that the
observed MDF of the NGC 5128 GCs can be reproduced accurately by a
formation scenario where roughly 65-75 percent of the mass in the
GC system was formed in an initial starburst 11-12 Gyrs ago, followed by a
second late starburst which peaked 2-4 Gyrs ago and produced the
remaining 25-35 percent of GC mass.

\subsection{Perturbations to the double starburst case}
As mentioned above, the $n$ starburst scenario contains $3n$ free
parameters. Clearly, as we make more free parameters available to
the model, we should expect to do better in terms of matching the
observed NGC 5128 MDF. We note first that the KS values from the
double starburst scenarios are significantly large. Refining the
model further would not make an appreciable difference in the KS
probabilities, certainly not as much as in the transition from the
single to the double starburst case. In addition, given the errors
in the observed MDF and the resolution of our model, further
refinement would not provide much greater insight into the GC
formation history of this galaxy.

Therefore, instead of repeating the study above with a wholesale
three starburst analysis, we chose to treat the third starburst
as a perturbation on the double starburst system. We were
interested in studying how the KS probabilities change if we apply
a small perturbation to each of the major star formation episodes.
This simple perturbation analysis gave an indication of how
localised these double starburst KS hotspots were in the nine
parameter triple starburst space.

\begin{figure}
\begin{center}
\includegraphics[width=3.4in]{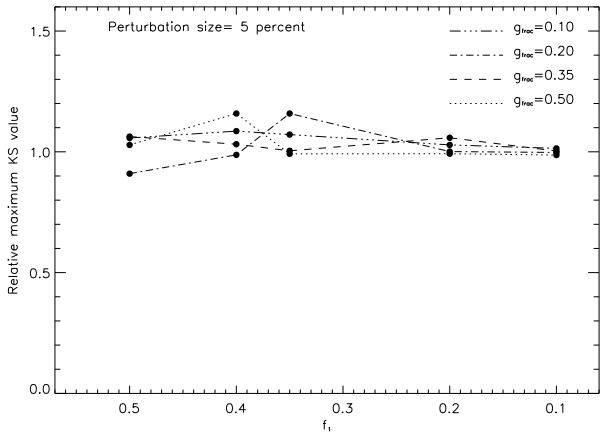}
\caption[KS values for a perturbed double starburst
scenario]{\small{KS values for perturbed double starburst
scenarios as a fraction of the KS values in the unperturbed case.
The size of the perturbation applied to the system is 5 percent.}}
\label{fig:perturbation_analysis}
\end{center}
\end{figure}

Fig. \ref{fig:perturbation_analysis} shows the best KS values
achieved after applying a 5 percent pertubation i.e.
$f_{perturb}=0.05$ as a \emph{fraction} of the best KS values in
the corresponding unperturbed double starburst scenario. In almost
all cases the KS value improves only marginally. We conclude from
this analysis that the KS hotspot is \emph{not} extremely
localised in the nine parameter triple starburst space and does appear
to be stable at least under perturbations of sizes 5 percent and
below. Although we have sampled only a small fraction of this
nine-dimensional parameter space, the resistance to small
perturbations indicates that the double burst scenarios are
reasonably stable and robust solutions.

\section{Discussion and conclusions}
We studied the formation history of the NGC 5128 GC system by
using a chemical enrichment model to accurately reproduce its
observed MDF derived from recently obtained $U$ and $B$ photometry.
Our results show that the GC system in this galaxy could be the
product of two major GC formation episodes, although we do not
have adequate resolution in the model or in the observed MDF to
make any reliable claim about the possibility of additional
formation events.

The double starburst analysis produces high KS probabilities and
therefore good fits to the observed MDF, with best-fit models
(shaded region in Fig. \ref{fig:relative_KS_values}) that favour an
initial GC formation episode 11-12 Gyrs ago which produced 65-75
percent of mass in the GC system, with a second late
formation episode approximately 2-4 Gyrs ago producing the
remaining 25-35 percent. The late starburst results in a small
fraction ($\sim$ 5 percent) of the stellar mass in the GC system
potentially having ages \emph{less than 1 Gyr}. If NGC 5128 has over 1500
clusters, then we might expect a handful of those clusters to have
ages less than 1 Gyr. We have not made any \emph{a priori}
assumptions about the driving mechanisms behind star formation
episodes in our model. While a single starburst followed by
passive evolution can clearly be discounted, the nature of our
model admits any scenario where multiple episodes of GC
formation are possible.

Our results are in general agreement with \citet{Beasley2003}, who
obtain good fits to the observed NGC 5128 MDF using a
semi-analytical galaxy formation model \emph{with metal-poor GC
formation halted at $z \sim 5$}. A quantitative similarity between
this study and that of \citet{Beasley2003} is that the initial GC
formation episode (first gaussian starburst) in our
model, which gave rise to the metal-poor GC subpopulation, decays
rapidly to virtually zero (see Fig. \ref{fig:best_sfr})
within $\sim 2$ Gyrs of the formation of the galaxy; i.e. there is
\emph{effective truncation} of metal-poor
GC production at high redshift ($z \sim 3-4$). This closely resembles
the truncation employed by \citet{Beasley2003}, and the similarity is
probably due to the fact that our model is comparable to the chemical enrichment
prescription employed in the semi-analytical model of
\citet{Beasley2003}. Since the object of both studies is to reproduce \emph{metallicity
distributions} this resemblance is not unexpected.

This study would not be complete without a comparison of our results
to those of \citet{Peng2004b}, who performed a spectroscopic
analysis of the GC dataset used in this paper. Based
on the age-sensitive $H_{\beta}$ index, they conclude that metal-poor
GCs in NGC 5128 have ages comparable to those in the
Milky Way, i.e. $\sim 12$ Gyrs \citep[e.g][]{Krauss2003}, and that
the metal-rich GCs are consistent with a mean age of $5^{+3}_{-2}$
Gyrs. We thus find that our age estimates, derived using a
chemical enrichment approach to exploit metallicity-sensitive
photometric colours, are consistent with a study of the same objects
using age-sensitive spectroscopic indices.

Our study demonstrates the potential of a chemical enrichment approach
in deciphering the formation histories of the GC system in galaxies.
As more age and metallicity-sensitive spectro-photometric data become
available, methods
such as the one used in this study will enable us to set robust
constraints on the way galaxies either form or incorporate GCs,
crucial to our understanding of galaxy formation.

\begin{acknowledgements}
We are grateful to the referee for numerous comments and suggestions that
improved the quality of this manuscript. We are indebted to Eric Peng
for providing the NGC 5128 data on which this
work is based, prior to publication. We warmly thank Andr\'es
Jord\'an for his very careful reading of this manuscript and numerous
useful comments. We also thank Roger Davies, Joseph Silk and
Julien Devriendt for constructive remarks regarding this study. SK acknowledges PPARC grant
PPA/S/S/2002/03532. This research was supported by PPARC
Theoretical Cosmology Rolling Grant PPA/G/O/2001/00016 (S. K. Yi
and I. Ferreras), the Glasstone Fellowship and the
 Post-doctoral Fellowship Program of Korea Science \& Engineering
 Foundation (S.-J. Yoon) and made use of
Starlink computing facilities at the University of Oxford.
\end{acknowledgements}

\bibliographystyle{aa}
\bibliography{references}

\end{document}